\documentstyle[aps,prd,preprint,tighten,floats,epsf]{revtex}

\def \pom {{\hspace{ -0.1em}I\hspace{-0.2em}P}}

\begin{document}
%

\preprint{PSU/TH/169}

\title{DIFFRACTIVE FACTORIZATION \ - \ A SIMPLE FIELD THEORY
MODEL FOR $F_2^{diff}(\beta x_{\pom}, Q^2; x_{\pom},t)$}
\author{ Arjun Berera}

\address{
   Department of Physics,
   Pennsylvania State University,
   University Park, PA 16802, U.S.A.
}

\maketitle

\begin{abstract}
Operator definitions of diffractive parton distribution functions
are given.  A distinction is made between the special case of ``Regge
factorization'' to the general case of ``diffractive factorization''
with explicit expressions for $F_2^{diff}(\beta x_{\pom}, Q^2;x_{\pom},t)$
in both cases.
A calculation from a simple field theory model is presented in the
style of ``constituent counting rules'' for the behavior of the diffractive
parton distribution
functions when $\beta \rightarrow 1$, which
corresponds to when the detected parton carries almost all of the
longitudinal momentum transferred from the scattered hadron.
A comment is made about the consistency of the model
with the observed flattening of $n(\beta)$ as $\beta \rightarrow 1$,
which recently was reported by the H1-collaboration
from their preliminary 1994 data.
\end{abstract}

\medskip

To appear in Proceedings of the International Workshop
on Deep-Inelastic Scattering and Related Topics, Rome, Italy
1996, ed. G. D' Agostini

\eject



In this talk I will discuss factorization in diffractive DIS.  In the general
phenomena of diffractive hard scattering, the initial proton in DIS
or even both protons at hadron collider participate in a hard process involving
a very large momentum transfer, but one or at hadron colliders one or both
hadrons
is diffractively scattered, emerging
with a small transverse momentum and the loss of a rather
small fraction of longitudinal momentum.

As shown by CFS \cite{cfs}, hard factorization breaks down
at leading twist for pure hadronically initiated hard diffraction
processes.  This is discussed further in my talk on double pomeron
exchange \cite{abtalk} presented at this conference.
However, at HERA we can hypothesize factorization
for diffractive DIS.

In the first stage we hypothesize that the
diffractive structure function $F_2^{\rm diff}$ can be written in
terms of a diffractive parton distribution :
\begin{equation}
{d F_2^{\rm diff}(\beta x_{\pom},Q^2;x_\pom,t)\over dx_\pom\,dt}
=x_{\pom} \sum_a \int_{\beta}^{1} d\beta '\
{d\, f^{\rm diff}_{a/A}(\beta ' x_{\pom},\mu;x_\pom,t)\over dx_\pom\,dt}\
\hat F_{2,a}(\beta /\beta ',Q^2;\mu),
\label{factor}
\end{equation}
where $\hat F_2$ is the same function which is convoluted with the
inclusive parton densities to compute $F_2$ of inclusive DIS.
If for simplicity, we ignore $Z$ exchange, then
%
%
$\hat F_{2,a}(\beta/\beta ',Q^2;\mu)
= e_a^2\ \delta(1-\beta/\beta ') + {\cal O}(\alpha_s)$.
%
%

In the second stage, we hypothesize that the diffractive parton distribution
function has a particular form:
\begin{equation}
{d\,f^{\rm diff}_{a/A}(\beta x_{\pom},\mu;x_\pom,t)\over dx_\pom\,dt}
={1 \over 8\pi^2}\, |\beta_{A}(t)|^2 x_\pom^{-2\alpha(t)}\,
f_{a/\pom}(\beta, t,\mu)\,.
\label{pomdist}
\end{equation}
Here $\beta_A(t)$ is the pomeron coupling to hadron A and $\alpha(t)$
is the pomeron trajectory.
The function $f_{a/\pom}(\beta, t,\mu)$ defined above is the
``distribution of partons in the pomeron''.
I distinguish the ``diffractive factorization'' of
Eq.~(\ref{factor}) from the ``Regge factorization''
of Eq.~(\ref{pomdist}).
The latter is a special case of the former.
The Ingelman-Schlein model \cite{IS} is synonymous with "Regge
factorization''.  The structure function
$F_2^{diff}(\beta x_{\pom}, Q^2; x_{\pom}, t)$ for the IS-model
is obtained by inserting
Eq.~(\ref{pomdist}) into
(\ref{factor}).  An inconsistency of data to the IS-model does not
also imply an inconsistency to diffractive factorization.

I now give operator definitions of the diffractive
parton distribution. The diffractive distribution of a quark of
type $j \in \{u,\bar u,d,\bar d,\dots\}$ in a hadron of type $A$
in terms of field operators $\tilde\psi(y^+,y^-,{\bf y})$
evaluated at $y^+ = 0$, ${\bf y} =0$ is:
\begin{eqnarray}
&&{d\, f^{\rm diff}_{a/A}(\beta x_{\pom},\mu; x_{\pom}, t)\over
dx_\pom\,dt}  =
{1 \over 64{\pi}^3}{1 \over 2}\sum_{s_{\!A}}\int d y^-
e^{-i\beta x_{\pom} P_{\!A}^+ y^-}
\nonumber\\
&& \sum_{X,s_{\!A^\prime}}
\langle P_{\!A},s_{\!A} |\tilde {\overline\psi}_j(0,y^-,{\bf 0})
| P_{\!A^\prime},s_{\!A^\prime}; X \rangle
\gamma^+ \langle P_{\!A^\prime},s_{\!A^\prime}; X|
{\tilde {\psi}}_j(0)| P_{\!A},s_{\!A} \rangle.
\label{fdiff1}
\end{eqnarray}
We sum over the spin $s_{\!A^\prime}$ of the final state proton and over
the states $X$ of any other particles that may accompany it.
Similarly, the diffractive distribution of gluons in a hadron is
\begin{eqnarray}
&&{d\, f^{\rm diff}_{a/A}(\beta x_{\pom},\mu; x_{\pom}, t)\over
dx_\pom\,dt} =
{1 \over 32{\pi}^3 \beta x_{\pom} P_{\!A}^+}{1 \over 2}\sum_{s_{\!A}} \int d
y^-
e^{-i\beta x_{\pom} P_{\!A}^+ y^-}
\nonumber\\
&&\sum_{X,s_{\!A^\prime}}
\langle P_{\!A},s_{\!A} |\tilde F_a(0,y^-,{\bf 0})^{+\nu}
| P_{\!A^\prime},s_{\!A^\prime}; X \rangle
\langle P_{\!A^\prime},s_{\!A^\prime}; X|
\tilde F_a(0)_\nu^{\ +}| P_{\!A},s_{\!A} \rangle.
\label{fdiff2}
\end{eqnarray}
The proton state $| P_{\!A},s_{\!A} \rangle$ has spin $s_{\!A}$ and
momentum $P_{\!A}^\mu =  (P_{\!A}^+, {M_{\!A}^2 / [2P_{\!A}^+]},{\bf
0})$. We average over the spin. Our states are normalized to
%
%
$\langle k |p \rangle =
(2\pi)^3\, 2p^+\,\delta(p^+ - k^+)\,\delta^2({\bf p} - {\bf k})$.
%
%
The tilde on the fields $\tilde \psi_j(0,y^-,{\bf 0})$
and $\tilde F_a(0,y^-,{\bf 0})^{+\nu}$ is to imply that they are
multiplied by an exponential of a line
integral of the vector potential as shown in \cite{bersop2}.

The diffractive parton
distributions are ultraviolet divergent and require
renormalization. It is convenient to perform the renormalization
using the $\overline{\rm MS}$ prescription, as
discussed in \cite{CSdistfns,CFP}. This introduces a renormalization
scale $\mu$ into the functions. In applications, one sets $\mu$ to be
the same order of magnitude as the hard scale of the physical process.

The renormalization involves ultraviolet divergent subgraphs.
Subgraphs with more than two
external parton legs carrying physical polarization
do not have an overall divergence.
Thus the divergent subgraphs are the same as for the ordinary parton
distributions. We conclude that the renormalization group equation
for the diffractive parton distributions is
\begin{equation}
\mu { d  \over d\mu}\,
{ d f^{\rm diff}_{a/A}(\beta x_{\pom},\mu;x_\pom,t) \over dx_\pom\,dt}=
\sum_b \int_{\beta x_{\pom}}^1 { d{z} \over {z}}\
P_{a/b}(\beta x_{\pom}/ z,\alpha_s(\mu))\
{ d f^{\rm diff}_{b/A}(z,\mu;x_\pom,t) \over dx_\pom\,dt}
\label{APeqn}
\end{equation}
with the same DGLAP kernel \cite{DGLAP},
$P_{a/b}( \beta x_{\pom} /z,\alpha_s(\mu))$, as one uses for the evolution of
ordinary parton distribution functions.

The diffractive parton distribution ${d\,f^{\rm diff}_{a/A}
(\beta x_{\pom},\mu;x_\pom,t)/ dx_\pom\,dt}$, like the ordinary parton
distribution, is essentially not calculable using perturbative
methods. Recall, however, that it is possible to derive ``constituent
counting rules'' that give predictions for ordinary parton
distributions ${f_{a/A}(x,\mu)}$ in the limit $x \to 1$ for not too
large values of the scale parameter $\mu$ in the sense
of the analysis by Brodsky and Farrar \cite{brofar}. In the
same spirit, in \cite{bersop2} we have considered the diffractive
parton distributions in the limit $\beta \to 1$.

In our
model the "pomeron" is represented by a 2-gluon exchange.  Inherently
the pomeron involves soft physics and its dynamics are unknown from QCD.
However 2-gluon models have been successful in describing some aspects
of the hard physics in diffractive hard processes.
Within the context of our model, we find in the limit
$\beta \rightarrow 1$, that there is an exact separation between the
hard partonic physics, which is measured, and the soft pomeron
(or better stated colorless exchange) physics, which is required
in order for the proton to diffractively scatter into the final state.
In spacetime the interpretation is this kinematic limit forces
the entire ``pomeron'' to be probed as a pointlike object.

We find that the diffractive gluon distribution behaves as
$ (1-\beta)^p$
%
%
%
%
for $\beta \to 1$ at moderate values of the scale $\mu$, say 2
GeV, with
$0 \leq p \leq 1$.
%
%
%
%
The choice $p\approx 0$ corresponds to an effectively massless final
state gluon, while $p\approx 1$ corresponds to an effective gluon
mass.
For the diffractive quark distribution we find they behave as
$(1-\beta)^2$.
%
%
%
%
However, suppose that we interpret the calculation
as saying that the diffractive distribution of gluons is
proportional to $(1-\beta)^0$ for $\beta$ near 1 when the scale $\mu$ is
not
too large. Then the evolution equation for the diffractive parton
distributions will give a quark distribution that behaves like
\begin{equation}
{df_{q/A}^{\rm diff}(\beta x_\pom,\mu;x_\pom,t) \over {dx_\pom\,dt}}
\propto (1-\beta)^1,
\end{equation}
when the scale $\mu$ is large enough that some gluon to quark
evolution has occurred, but not so large that effective power $p$ in
$(1-\beta)^p$ for the gluon distribution has evolved substantially
from $p = 0$. A signature of this phenomenon is that the diffractive
quark distribution will be growing as $\mu$ increases at large
$\beta$, rather than shrinking. Perhaps this is seen in the data
\cite{ZeusH1}.

{\em Post Conference Comment} $:$  Let us examine
the $\beta$-dependence of the pomeron
intercept, as reported by the H1-collaboration at this
conference from their 1994 preliminary data \cite{new}.
To clarify conventions, the parameterization and notation used
by H1 for the diffractive structure function is
\begin{equation}
F_2^D = A(\beta, Q^2) \frac{1}{x_{\pom}^n}.
\end{equation}
H1 is reporting that $n$ depends on $\beta$, so more appropriately
$n(\beta)$.

$\beta$ is a kinematic variable associated with the hard physics.
I know of no theoretical argument that precludes $\beta$ from
affecting the soft physics for general value of $\beta$.
However the hard/soft separation found in our model
(discussed after eq. (5)) as $\beta \rightarrow 1$
suggests that the $\beta$ dependence in the soft physics should
diminish in this limit.  It is therefore reassuring to see
from the 94 H1-preliminary data that the measured
curve for $n(\beta)$ flattens as $\beta \rightarrow 1$.

I clarify that a $\beta$-dependence in the intercept does not imply
a breakdown of diffractive factorization.  For this the intercept
(or equivalently $n$) needs a $Q^2$-dependence.	 which is not
found in Zeus `93
and up to H1 `94 data \cite{ZeusH1,new}.

\section*{Acknowledgments}

I thank R. Ball and V. Del Duca for their invitation.
This work was supported by the U.S. Department
of Energy.


\begin{references}


\bibitem{cfs} J.C. Collins, L. Frankfurt, and M. Strikman,
{\em Phys.\ Lett.} B {\bf 307}, 161 (1993).

\bibitem{abtalk} A. \ Berera, these proceedings.

\bibitem{IS}  G.\ Ingelman and P.\ Schlein,
{\em Phys.\ Lett.} B {\bf 152}, 256 (1985).

\bibitem{bersop2} A.\ Berera and D.\ E.\ Soper,
"Behavior of Diffractive Parton Distribution Functions", hep-ph/9509239,
(in press {\em  Phys.\ Rev.} D 1996).

\bibitem{CSdistfns} J.\ C.\ Collins and D.\ E.\ Soper,
{\em Nucl.\ Phys.}  B {\bf 194}, 445 (1982).

\bibitem{CFP}
G.\ Curci, W.\ Furmanski and R.\ Petronzio,
{\em Nucl.\ Phys.} B {\bf 175}, 27 (1980).

\bibitem{DGLAP} V.\ N.\ Gribov and L.\ N.\ Lipatov,
{\em Sov.\ J.\ Nucl.\ Phys.} {\bf 15}, 438 (1972);
Yu.\ L.\ Dokshitzer, {\em Sov. Phys. JEPT} {\bf 56}, 641 (1977);
G.\  Altarelli and G.\  Parisi, {\em Nucl.\ Phys.} B {\bf 26}, 298 (1978).

\bibitem{brofar}
S.\ J.\ Brodsky and G.\ Farrar,
{\em Phys.\ Rev.} D {\bf 11}, 1309 (1975).

\bibitem{ZeusH1} ZEUS Collaboration
(M. Derrick, {\it et al.}), {\em Z. Phys.} C {\bf 70}, 391 (1996);
H1 Collaboration, (T.\ Ahmed {\it et al.}),
{\em Phys.\ Lett.} B {\bf 348}, 681 (1995).


\bibitem{new} P. \ Newman (H1 Collaboration), these proceedings.

\end{references}
\end{document}